\documentclass[aps,pra,twocolumn,showpacs,showkeys]{revtex4-1}
\usepackage[T2A,OT1]{fontenc}
\usepackage[cp1251]{inputenc}         
\usepackage[english]{babel} 
\usepackage[dvips]{epsfig}            
\usepackage{dcolumn}
\usepackage{latexsym}
\usepackage{amsfonts}
\usepackage{xcolor}[2007/01/21]
\usepackage[bookmarks=true,bookmarksnumbered,bookmarksopen]{hyperref}
\hypersetup{pdfpagemode=FullScreen,colorlinks=true, citecolor=green,unicode,bookmarksopenlevel=3,
pdftoolbar=true,pdfauthor={Helen Gomonay},pdftitle={AFM/NM/FM}}%
\begin{document}
\title{Magnetostriction-induced anisotropy in the exchange biased bilayers}
\author{Olena Gomonay}
\affiliation {
 National Technical University of Ukraine ``KPI''\\ ave Peremogy, 37, 03056, Kyiv,
Ukraine}
 \author{Igor Lukyanchuk}
 \affiliation{Laboratory of Cond. Mat. Physics, University of Picardy, LPMC-UPJV, 33 rue Saint-Leu 80039 Amiens, FRANCE}
 \date{\today}
 \begin{abstract}
The exchange bias at ferromagnetic (FM)/antiferromagnetic (AF) interfaces strongly depends upon the state of antiferromagnetic layer which, due to strong magnetoelastic coupling, is sensitive to mechanical stresses. In the present paper we consider magnetoelastic effects that arise at FM/AF interface due to lattice misfit and magnetic ordering. We show how magnetostriction affects mutual orientation of AF and FM vectors and easy-axis direction in thin AF layer. The results obtained could be used for tailoring exchange biased systems.
 \end{abstract}
   \maketitle

\section{Introduction} 
Antiferromagentic (AF) materials are widely used in spintronic devices as auxiliary elements for pinning of ferromagnetic (FM) magnetization through the effect of exchange bias (see, e.g.\cite{Li:2007:1533-4880:13}). The possibility to control the state of coupled AF/FM bilayers requires investigation of the magnetic mechanisms that could be responsible for bias effect. Many researchers \cite{Radu:2009PhRvB..79r4425R,Ijiri:1998, Nogues:1999,
Miltenyi:2000, Zhu:2001, Nikitenko:PhysRevB.68.014418} emphasize the important role of the AF
domain structure in the establishing of the exchange bias. The
problem of AF domains is intimately related with magnetoelastic
coupling \cite{gomo:PhysRevB.75.174439} and can strongly depend upon the mechanical stress that appears at the FM/AF interface due to the lattice misfit. Magnetostriction can also provide additional coupling between FM and AF layers and affect orientation of AF moments in the near-surface region  \cite{Scholl:PhysRevB.84.220410, Scholl:doi10.1021/nl060615f}. Widely studied
epitaxial films consisting of FM and nonmagnetic materials
\cite{Sander:1999, Enders:1999, Gutajahr:2000,Wulfhekel:2001}
show strong correlation between magnetoelastic coupling and magnetic properties. Analogous and even more striking phenomena could be expected in the systems which combine FM with AF that posseses large magnetostriction.

In the present paper we show that magnetostriction of AF produces uniaxial anisotropy in the plane of the adjacent FM layer and thus causes strong surface magnetic anisotropy in AF itself.

\section{Uniaxial anisotropy of ferromagnet}
Epitaxial ferromagnetic film deposited on top of AF inherits the crystallographic structure of the substrate. If the substrate has a certain anisotropy induced by magnetoelastic strains, then, this anisotropy in atomic arrangement will be reproduced by the FM layer. Thus, additional contribution to the magnetic energy of the film should be proportional to magnetoelastic coupling in both FM and AF materials. Phenomenological expression for a such type of uniaxial in-plane anisotropy can be deduced from the magnetoelastic energy of FM which for a cubic-symmetry crystal follows as
\begin{eqnarray}\label{1}
 f^{\rm F}_{\rm me}&=&
b^{\rm F}_1[u_{xx}\alpha^2_x+ u_{yy}\alpha^2_y+u_{zz}\alpha^2_z]\nonumber\\
&+&2b^{\rm
F}_2[\alpha_x\alpha_yu_{xy}+\alpha_y\alpha_zu_{yz}+\alpha_z\alpha_xu_{zx}].
\end{eqnarray}
Here $u_{ik}$ are strain tensor components which we calculate with respect to the bulk nonmagnetic reference state, $b^{\rm F}_{1,2}$
 are magnetoelastic coupling coefficients. Magnetisation vector ${\mathbf M}_{\rm F}$ of
 FM is described
by the direction cosines $\alpha_k$, $k = x, y, z$. In the
relaxed state of FM/AF system an equilibrium strain $u_{ik}$
 includes deformations
produced by lattice mismatch $\epsilon_{\rm MF}$ and spontaneous
strain $\hat u_{\rm mag}$ induced by magnetic ordering in the AF
substrate. For a symmetric (001) surface the misfit-induced strains are isotropic and can influence only out-of-plane anisotropy of FM. In contrast, magnetostrictive contribution, though small as compared with the misfit strain, has nontrivial shear components,  $u^{\rm AF }_{xx}-u^{\rm AF }_{yy}$ and/or
$u^{\rm AF}_{xy}$ which can remove degeneracy between different
in-plane directions. Thus, uniaxial contribution into
magnetocrystalline energy of FM film takes a form
\begin{equation}\label{2}
 f^{\rm F}_{\rm ua} ={1 \over 2}K^{\rm F}_{1\mathrm
 ua}\rho_j(\alpha^2_x-\alpha^2_y)+K^{\rm F}_{2\mathrm
 ua}\rho_j\alpha_x\alpha_y,
\end{equation}
with anisotropy constants
\begin{equation}\label{uniax}
  K^{\rm F}_{1\mathrm ua}=b^{\rm F}_1(u^{\rm AF
}_{xx}-u^{\rm AF }_{yy}),\qquad K^{\rm F}_{2\mathrm ua}=2b^{\rm
F}_2u^{\rm AF}_{xy}.
\end{equation}
Variable $\rho_j=\pm 1$ distinguishes between the different AF domains.

A preferable direction of FM magnetisation ${\mathbf M}_{\rm F}$,
which depends upon the sign of the coefficients $K^{\rm F}_{\rm
ua}$, is defined by correlation between self-striction of the FM
and external striction imposed by the AF. If, for example,
magnetostriction of FM in the direction of magnetisation is
positive (elongation, $b^{\rm F}<0$), then, ${\mathbf M}_{\rm F}$
will tend to align in the direction of maximal elongation of the
AF, i.e., for positive $u^{\rm AF}$ value (elongation) $K^{\rm
F}_{\rm ua}$ is negative, as can be easily checked from equation
(\ref{uniax}).

Magnetostriction-induced uniaxial anisotropy (\ref{uniax})
competes with the anisotropy arising from the FM/AF exchange in a
thin near-surface region of thickness $\xi$. For a compensated AF
surface this contribution depends upon the exchange integral
between the atoms of F and AF $J_{\rm F-AF}$ and susceptibility of
AF $\chi_{\rm AF}\equiv 1/J_{\rm AF}$ (Koon's model, \cite{Koon:1997}):
\begin{equation}\label{12}
f_{\rm exch}=-{1\over 2}\chi_{\rm AF}J^2_{\rm F-AF}[{\mathbf
M}_{\rm  F}\times {\mathbf L}_S]^2,
\end{equation}
The AF vector ${\mathbf L}_S$ describes orientation of spins at
the surface of the AF substrate (which in principle can differ
from that in the bulk, as will be shown later).

To elucidate the effect of both contributions let us consider a
simple case when one of the in-plane easy axis (say, $x$) of FM
coincides with in-plane ${\mathbf L}_S$ direction and $u^{\rm
AF}_{xy}=0$. For the in-plane FM ordering ($\alpha_z=0$) we set
$\alpha_x=\cos\psi$, $\alpha_y=\sin\psi$. The effective energy is
thus
\begin{equation}\label{14}
 f^{\rm F}_{\rm eff}={1 \over 4}K_4\sin^22\psi+{1 \over 2}[K_{\rm ua}\rho_j+\frac{\xi}{2t_{\rm F}}\chi_{\rm
AF}J^2_{\rm F-AF}]\cos 2\psi.
\end{equation}
Constant $K_4>0$ is magnetocrystalline constant, and we suppose
the FM film to be homogeneously ordered throughout the thickness
$t_{\rm F}$.

Equilibrium value $\psi=\psi^\mathrm{eq}$  minimizes effective energy (\ref{14}), so, it satisfies the relations 
\begin{eqnarray}\label{15}
&&\sin2\psi^\mathrm{eq}\{K_4\cos 2\psi^\mathrm{eq} -[K_{\rm ua}\rho_j+\frac{\xi}{2t_{\rm
F}}\chi_{\rm
AF}J^2_{\rm F-AF} ]\}=0\nonumber\\
&&K_4\cos 4\psi^\mathrm{eq} -[2K_{\rm ua}\rho_j+\frac{\xi}{t_{\rm F}}\chi_{\rm
AF}J^2_{\rm F-AF} ]\cos 2\psi^\mathrm{eq}
>0.
\end{eqnarray}

In the absence of FM/AF interaction, FM has two equivalent easy
directions in (001) plane, $\psi^{(0)}_1=0$ and $\psi^{(0)}_2=\pi/2$. Antiferromagnetic substrate removes this degeneracy.  If exchange coupling
is not too large, $\xi\chi_{\rm AF}J^2_{\rm F-AF}\leq K_4t_{\rm
F}$, both  solutions $\psi^\mathrm{eq}_{1,2}$ satisfy equations (\ref{15}),
but have different energies, the difference being
\begin{equation}\label{4}
f^{\rm F}_{\rm eff}(\psi^\mathrm{eq}_1)-f^{\rm F}_{\rm eff}(\psi^\mathrm{eq}_2)=K_{\rm
ua}\rho_j+\frac{\xi}{2t_{\rm F}}\chi_{\rm AF}J^2_{\rm F-AF}.
\end{equation}
It can be easily seen from (\ref{4}) that the FM/AF exchange coupling makes favourable the solution with ${\mathbf M}_{\rm F}\perp {\mathbf L}_S$
($\psi^\mathrm{eq}_2=\pi/2$) for any sign of the exchange constant  $J_{\rm F-AF}$. In turn, magnetostriction-induced anisotropy
$K_{\rm ua}$ may oppose this tendency and make preferable
in-parallel orientation of ${\mathbf M}_{\rm F}$ and ${\mathbf
L}_S$ ($\psi^\mathrm{eq}_1=0$). It should be stressed that these two
mechanisms have different origin and the system can switch from
one easy-axis to another with variation of FM thickness. Exchange
mechanism ties together mutual orientation of FM magnetisation and
AF spins in the near-surface layer. This mechanism is important
for very thin films where factor $\xi/t_{\rm F}$ is not
vanishingly small. Magnetostriction-related mechanism is a
long-range one, it depends upon orientation of AF moments in the
bulk which can be different from ${\mathbf L}_S$. Moreover, in
some AFs widely used in FM/AF systems (e.g., NiO, CoO, LaFeO$_3$,
KCoF$_3$) magnetostriction originates from the strong spatial
dependence of the exchange integral and is insensitive to exact
orientation of AF spins. In this very important case uniaxial
anisotropy of FM is defined mainly by the domain structure of AF.

The role of magnetostriction-induced mechanism can be illustrated
by some experimental examples. Simultaneous observation of the FM
and AF spins in Co/LaFeO$_3$ \cite{Nolting:2000} and Co/NiO
\cite{Ohldag:2001} systems revealed that FM magnetisation is
aligned parallel or aniparallel to the in-plane projection of the
AF axis in contrast to the usually observed perpendicular
coupling consistent with the Koon's model \cite{Koon:1997}.
Uniaxial anisotropy was also detected after deposition of Fe on
top of KCoF$_3$ \cite{Malkinski:2002(2), Malkinski:2002}. All these AFs are known to have rather large magnetostriction of the exchange nature (see Table \ref{Tab:AF})
.
\begin{table}
  \centering
  \caption{Magnetostriction (spontaneous deformations) of typical AFs calculated the experimentally observed lattice consants above and below Neel temperature}\label{Tab:AF}
  \begin{tabular}{ccc}
\hline AF&Magnetostriction&\\ \hline
    NiO & -2.6$\cdot 10^{-3}$\cite{Yamada:1966} &  \\
    LaFeO$_3$ & -4.76$\cdot 10^{-4}$ \cite{Abrahams:1972} &  \\
     KCoF$_3$ & -2.0$\cdot 10^{-3}$\cite{Julliard:1975, Skrzypek:1995}  &  \\
     CoO & -2$\cdot 10^{-2}$\cite{Greiner:1966} &  \\\hline
  \end{tabular}
\end{table}
Using the values of magnetoelastic constants for ferromagnets (Table \ref{Tab:FM}), one can calculate from equation (\ref{uniax}) the expected value of uniaxial anisotropy in different FM/AF combinations (see Table \ref{Tab:FM/AF}). 

\begin{table*}
  \centering
  \caption{Magnetoelastic coupling coefficients for FMs \cite{Sander:1999}, in  erg/cm$^3$}\label{Tab:FM}
  \begin{tabular}{ccc}
\hline &Co, fcc&Fe, bcc\\\hline
 $ b_1$&-9.2$\cdot 10^{7}$&-3.43$\cdot 10^{7}$
\\
 $ b_2$&7.7$\cdot
10^{7}$&7.83$\cdot 10^{7}$
\\\hline
  \end{tabular}
\end{table*}
As can be seen from Table \ref{Tab:FM/AF}, uniaxial anisotropy in
the Fe film constitutes only 10 $\%$ from the ``pure'' magnetic
anisotropy. Nevertheless, this value can be enough to chose
preferable axis of magnetisation as was clearly observed in the
experiments \cite{Malkinski:2002(2),Malkinski:2002}.  More pronounced effect is expected in Co films which have rather high magnetostriction and small bulk magnetic anisotropy. Predicted value of the uniaxial anisotropy is of the same order as $K_4$ or even one order of magnitude larger, as in the case of Co/CoO. It should be noted, that in calculation we started from the bulk values of magnetoelastic coefficients for Fe and Co. In the case of ultrathin Co films these values need to be ascertained because of the large potential misfit between FM and AF lattices (nearly
10 $\%$). Depending on the growth mode this mismatch can either relax through the formation of dislocations or produce strong internal stresses in the Co film which, in turn, can give rise to a crucial change of the value and even the sign of magnetoelastic coefficient  (see, e.g.\cite{Bochi:1994, Gutajahr:2000}).
\begin{table*}
 \begin{center}
  \caption{
Magnetic anisotropy of FM/AF systems, erg/cm3. $K_4$ (2nd column) is the 4-th order magnetocrystalline anisotropy observed in the bulk Fe and Co crystals. Theoretical values of $K_\mathrm{ua}$ (2nd column) are calculated from equation (\ref{uniax}). Experimental values of $K_\mathrm{ua}$ (the last column) are extracted from from measurement of hysteresis loops (for Co) and ferromagnetic resonance (for Fe). 
  }\label{Tab:FM/AF}
 \begin{tabular}{cccc}
\hline &$K_4$,
erg/cm$^3$&\multicolumn{2}{c}{$K_{ua}$, erg/cm$^3$}\\
&&theor&exper\\\hline
    Co/NiO & -2.3$\cdot 10^{5}$\cite{Landolt:1982(19)}& 2.0$\cdot 10^{5}$ &1.8$\cdot 10^{5}$\cite{Chopra:2000}\\
Co/LaFeO$_3$ &  & 0.37$\cdot 10^{5}$ &1.4 $\cdot 10^{5}$\cite{Nolting:2000}\\
Co/CoO&  & 6.0$\cdot 10^{6}$ &1.2$\cdot
10^{6}$\cite{Gredig:2002}
\\\hline
    Fe/NiO & 8.5$\cdot 10^{5}$\cite{Enders:1999,Malkinski:2002} & 0.9$\cdot 10^{5}$&-\\
     Fe/KCoF$_3$ &  &0.7$\cdot 10^{5}$  &0.8$\cdot 10^{5}$\cite{Malkinski:2002}\\\hline
  \end{tabular}
   \end{center}
\end{table*}

\section{Surface anisotropy of antiferromagnet}
It is widely recognised that lattice misfit strongly influences the magnetic and
magnetoelastic properties of the film (see,
e.g. \cite{Victora:1993}). On the other hand, epitaxial misfit
may equally induce large stress in the substrate (this phenomenon
is used to measure stress in the film \cite{Sander:1999}). In the
case when the substrate is rather thick, stress exerted by the
film relaxes over a small distance $\Sigma^{\rm AF}$ in the
near-surface layer of AF. For AFs with large magnetoelastic
coupling this  surface stress can produce an additional magnetic
anisotropy which we will call a surface anisotropy.

Phenomenological description of this effect is based on the
analysis of the Helmholtz free energy potential $G$ which
includes elastic $f_{\rm e}$ and magnetoelastic $f^{\rm AF}_{\rm
me}$ energy of AF using antiferromagnetic
vector ${\mathbf L}$ and components of stress tensor $\hat\sigma$ as the internal
  parameters:
\begin{equation}\label{3}
  G=\int_{AF}(f^{\rm AF}_{\rm me}+f_{\rm e})dV.
  \end{equation}
In the simplest case of a cubic crystal the elastic energy density $f_{\rm e}$ takes a form
\begin{eqnarray}\label{elastic}
f_{\rm e}&=&\frac{1}{2}s_{11}[\sigma_{xx}^2+\sigma_{yy}^2+\sigma_{zz}^2]+s_{12}[\sigma_{xx}\sigma_{yy}+\sigma_{yy}\sigma_{zz}+\sigma_{zz}\sigma_{xx}]\nonumber\\
&+&2s_{44}[\sigma^2_{xy}+\sigma^2_{yz}+\sigma^2_{zx}].
\end{eqnarray}
where we turned from strains to stresses using the Hook's law.
Compliances $s_{ik}$ are expressed through the elastic modula
$c_{ik}$ in a usual way:
\begin{eqnarray}\label{compliance}s_{11}&=&\frac{c_{11}+c_{12}}{(c_{11}-c_{12})(c_{11}+2c_{12})},
\nonumber\\
s_{12}&=&-\frac{c_{12}}{(c_{11}-c_{12})(c_{11}+2c_{12})},\nonumber\\
s_{44}&=&\frac{1}{c_{44}}.\end{eqnarray} 
Density of magnetoelastic energy $f_{\rm me}$ can be written as
\begin{eqnarray}\label{magnetoelastic}
  f^{\rm AF}_{\rm me}&=&f_{\rm exch}
+\frac{b^{\rm AF}_1}{c_{11}-c_{12}}[\sigma_{xx}L^2_x+ \sigma_{yy}L^2_y+\sigma_{zz}L^2_z]\nonumber\\
&+&2\frac{b^{\rm
AF}_2}{c_{44}}[L_xL_y\sigma_{xy}+L_yL_z\sigma_{yz}+L_zL_x\sigma_{zx}],
\end{eqnarray}
\noindent where $b^{\rm AF}_{1,2}$ are magnetoelastic
coupling coefficients of a cubic AF, and far from the N\'eel temperature AF vector can be normalised, so
$|{\mathbf L}|=1$. The first term in
(\ref{magnetoelastic}) describes a possible nonisomorphic
contribution which arises from the space dependence of the
exchange interactions described by a coefficient $B^{\rm
AF}_{0}$. It depends upon the specific type of AF, for example,
for a single-domain NiO it can be expressed as
\[f_{\rm exch}=\frac{B^{\rm AF}_{0}}{c_{44}}(\sigma_{xy}+\sigma_{yz}+\sigma_{zx}){\mathbf L}^2.\]
In the presence of the FM coverage, the AF substrate exerts a surface
stress $\hat\tau^{\rm AF}$ opposite to the surface stress in the
FM film $\hat\tau^{\rm F}$:
\begin{equation}\label{AF_stress}
\hat\tau^{\rm AF}\equiv \int \hat\sigma dz=-\hat\tau^{\rm F}.
\end{equation}
\noindent $z$ axis is directed along the film normal and
integration is over the AF thickness. For a (001) cubic surface
$\hat\tau^{\rm F}$ can be estimated from the misfit value
$\epsilon_{MF}$ as follows:
\begin{equation}\label{surface_stress}
\tau^{\rm F}_{xx}=\tau^{\rm F}_{yy}=t_{\rm
F}\left(c_{11}+c_{12}-\frac{c_{12}^2}{c_{11}}\right)\epsilon_{MF},
\end{equation}
where $t_{\rm F}$ is the FM film thickness.  Substituting
(\ref{AF_stress}) into free energy (\ref{3}) we obtain a
contribution from the FM/AF misfit as
\begin{equation}\label{7}
G_{\rm eff}= \int_S\frac{b^{\rm AF}_1}{(c_{11}-c_{12})}\tau^{\rm
F}L^2_zdS=\frac{1}{2}\int_SK^{\rm AF}_{\rm S}L^2_zdS,
\end{equation}
which could be associated with the surface/interface energy of AF.
Effective constant
\begin{equation}\label{8}
K^{\rm AF}_{\rm S}=2b^{\rm
AF}_1\epsilon_{MF}\left(1+2\frac{c_{12}}{c_{11}}\right)t_{\rm F}
\end{equation}
is proportional to the product of magnetoelastic coupling
coefficient of AF and misfit (or effective stress) in the FM
layer.

The sign of $K^{\rm AF}_{\rm S}$ and hence, the character of the
induced surface anisotropy, is defined by the relation between
the sign of AF spontaneous striction and that of external stress.
Suppose, FM lattice constant is smaller than that of AF
($\tau^{\rm F}>0$). Then, AF surface exerts a compressive stress.
According to the general Le-Chatelier principle, AF vector at the
surface will rotate in a way which reduces the external
influence. In the case of positive striction (AF spontaneously
elongates in spin direction) in-plane orientation of AF spins
will be preferable ($K^{\rm AF}_{\rm S}>0$). It worth to mention that the analogous, magnetoelastic, mechanism related with the rotation of AF moments in near-surface region is responsible for the shape-induced magnetic anisotropy in AF nanoparticles \cite{Gomonay2014125}.

\section{Discussion}
The misfit-induced surface anisotropy can produce a noticeable
rotation of AF spins in the vicinity of interface region. The
most pronounced effect can be expected for NiO, CoO, and
LaFeO$_3$ AFs in which the bulk AF vector makes some angle with
(001) surface. Particularly, in NiO and CoO the AF spins are
ordered in (111) planes (with small deflection in the case of CoO
\cite{Roth:1958, Slack:1960, Herrmann:1978}) in which they can be
easily rotated. An easy-axis is directed along $\langle
2\bar{1}\bar{1}\rangle$ in NiO and $\langle
3\bar{1}\bar{1}\rangle$ in CoO, thus, for a cleaved (001) surface
  AF moments have nonzero component perpendicular to the surface plane, as was observed for a NiO crystal
  \cite{Hillebrecht:2001(2),Scholl:2001(2)}.

Deposition of Fe and Co on NiO, and Fe$_3$O$_4$ and Co on CoO
produces compressive surface stress in AF (see Table
\ref{Tab:lat_const}). which gives the values of interatomic
distance for different FM and AF at (001) surface of fcc lattice
(2-nd column) calculated from the bulk lattice parameters (1st
column)).

\begin{table*}
  \begin{center}
  \caption{Bulk lattice parameters (2nd column) and calculated interatomic distances at (001) surface of fcc lattice (3d column) different FM and AF, in  \AA.}\label{Tab:lat_const}
\begin{tabular}{ccc} \hline
&Bulk &(001) surface\\ \hline
  Fe, bcc &2.866 \cite{Landolt:1982(19)}  &4.053 \\
  Co, fcc & 3.544 \cite{Landolt:1982(19)}& 3.544 \\
  Fe$_3$O$_4$& 8.398 \cite{Ijiri:1998}&4.199  \\
  NiO&4.177&4.177\\
  CoO&8.508\footnotemark[1] \cite{Ijiri:1998}&4.254\\ \hline
\end{tabular}\\
\end{center}
\footnotetext{Magnetic unit cell}
\end{table*}
For NiO and CoO the magnetoelastic constant $b^{\rm AF}_1$ is
positive (as deduced from the data \cite{Greiner:1966, Yamada:1966(2)}), so, as it follows from (\ref{7}), (\ref{8}), for
all the mentioned FM/AF combinations the preferable orientation
of AF vector ${\mathbf L}$ is in the interface (001) plane. A
compromise between the strong dipole-dipole anisotropy which
tends to keep AF moments close to (111) plane and strain-induced
surface anisotropy in (001) plane is the direction [1$\bar{1}$0].
So, depending upon the balance between the bulk magnetic
anisotropy and the induced surface anisotropy (\ref{8}), AF
moments may rotate from the bulk easy direction to [1$\bar{1}$0]
to a smaller or larger angle. The effect should be obviously
stronger for Co FM because of the large misfit value.

 Experimentally  this phenomenon was observed in \cite{Ohldag:2001}, where deposition of 2 nm Co
 film on the (001) surface of NiO resulted in the total reorientation of NiO
 spins to [1$\bar{1}$0] direction. An observed collinear alignment of Co
 and NiO spins in this system arises from both misfit-induced reorientation of AF moments and magnetistriction-induced uniaxial anisotropy in the FM layer.

An analogous effect was observed in the Fe$_3$O$_4$/CoO
multilayers \cite{Ijiri:1998} where an influence of the surface
stress is much more pronounced. In this system all of the AF Co
moments lie along [110] or [$\bar{1}$10] directions (depending on the AF domain type). This orientation does not vary with temperature, magnetic field and thickness of CoO layers.

Misfit-induced anisotropy of AF layer depends upon the internal stresses $\hat{\tau}^\mathrm{F}$ in the adjacent ferromagnet which could relax in the course of field cycling. Variation of stress, in turn,  affects the domain structure of AF.  So, magnetoelastic mechanism can explain training effect (irreversible changes in configuration of AF domains) frequently observed in bilayers with multidomain state of AF in the as-cast sample (see, e.g. \cite{Radu:2009PhRvB..79r4425R, Fu:JMP2011}).

In summary, we have studied the effect of magnetostriction on the
properties of FM/AF coupled system. Spontaneous striction which
appears in antiferromagnet due to AF ordering can cause uniaxial
in-plane anisotropy in the ferromagnetic film and set preferential
easy axis of FM either along with or perpendicular to the orientation
of AF vector. Competition between uniaxial anisotropy induced by
long-range magnetostriction and short-range exchange mechanism
results in different orientation of the FM easy-axis depending
on the thickness of FM layer. Lattice misfit between FM and AF is
a source of a magnetic surface anisotropy in AF substrate which
can cause rotation of AF moments in the near-surface region
compared with their bulk orientation. 

Authors acknowledge the financial support from the FP7-SIMTECH.
 




%

\end{document}